\documentclass[11pt]{article}
\usepackage[utf8]{inputenc}
\usepackage{epsfig}
\usepackage{amsmath,amsfonts,latexsym, amssymb}

\newtheorem{satz}{Theorem}[section]

\newtheorem{conclusion}[satz]{Conclusion}
\newtheorem{ob}[satz]{Observation}

\newcommand{\mbf}{\mathbf}

\newcommand{\tit}{\textit}

\newcommand{\R}{\mathbb{R}}

\begin{document}
\thispagestyle{empty}
\begin{center}
\vspace*{1.0cm}

{\LARGE{\bf About the Minimal Resolution of Space-Time Grains\\
in Experimental Quantum Gravity }} 

\vskip 1.5cm

{\large {\bf Manfred Requardt }} 

\vskip 0.5 cm 

Institut f\"ur Theoretische Physik \\ 
Universit\"at G\"ottingen \\ 
Friedrich-Hund-Platz 1 \\ 
37077 G\"ottingen \quad Germany\\
(E-mail: requardt@theorie.physik.uni-goettingen.de)

\end{center}

\vspace{0.5 cm}

\begin{abstract}
 We critically analyse and compare various recent thought experiments,
 performed by Amelino-Camelia, Ng et al., Baez et al., Adler et al.,
 and ourselves, concerning the (thought)experimental accessibility of
 the Planck scale by space-time measurements. We show that a closer
 inspection of the working of the measuring devices, by taking their
 microscopic quantum many-body nature in due account, leads to deeper
 insights concerning the extreme limits of the precision of space-time
 measurements. Among other things, we show how certain constraints
 like e.g. the Schwarzschild constraint can be circumvented and that
 quantum fluctuations being present in the measuring devices can be
 reduced by designing more intelligent measuring instruments.
 Consequences for various phenomenological quantum gravity models are
 discussed.

\end{abstract} \newpage
\setcounter{page}{1}
\section{Introduction}
According to general folklore, originally mostly based on simple
dimensional considerations or qualitative reasoning and later
supported by certain gedanken experiments combined with a couple of
fundamental assumptions (to mention but a few sources see
e.g. \cite{Wheeler}, \cite{Padmanaban}, or the discussions in
\cite{Garay},\cite{Adler2}), there existed widespread
agreement that fundamental lower limits to space-time measurements and
resolution are roughly given by the respective Planck values, e.g. the
Planck length $l_P=(G\hbar/c^3)^{1/2}$. One must however concede that
in practice it seems to be presently impossible to come near these
values in real experiments. Therefore, most of the work is rather of a
thought-experimental character. 

More recently it has been argued by various groups that these alluded
fundamental bounds are in fact much larger and are perhaps just at the
brink of becoming observable by using the most recent class of
gravity-wave interferometers, more specifically, by observing the
effects of (geometric) vacuum fluctuations in e.g. length
measurements. As far as we can see, this particular field started more
or less with the two papers \cite{Ng1} and \cite{Cam1}, a precursor,
having however a slightly different focus, being \cite{Diosi}.     

What is most puzzling is the claim of the authors that (lower bounds
to the) uncertainty of length measurement turn out to be proportional
to the square root or a simple fractional power of the length (or
distance), $l$, to be measured, i.e. the fundamental uncertainty in
length measurement (or rather the respective lower bound) seems to
increase with $l$! This is, at first glance, quite unusual and perhaps
even counterintuitive. As it would represent quite a departure from
(perhaps too naive?) general accepted wisdom if these arguments turn
out to be correct, it is of tantamount importance to scrutinize the
correctness of the arguments being advanced in favor of this
opinion. We note in passing that we are quite sympathetic in general
to such an enterprise of developing a, so to speak, semi-classical
quantum gravity phenomenology.

In contrast to earlier work, this more recent line of arguments is
based on a paper by Wigner and Salecker (\cite{Wigner}), in which
distance measurements in general relativity are analysed if effects of
the Heisenberg uncertainty relation are included (as a typical
ingredient of quantum theory). Postponing these more technical points
to the following section, we continue with a brief discussion of the
historical evolution of the field.

While papers on this topic, essentially repeating the original
arguments, continue to appear up to quite recent times (see
e.g. \cite{Cam2},\cite{Cam3},\cite{Ng2},\cite{Ng3},\cite{Ng4}), one
should note that there have been a couple of contributions which
provided arguments against the claimed inevitability of the
fundamental measurement limitations of Jack Ng, van Dam and
Amelino-Camelia, in particular concerning the strange dependence on
the (macroscopic) distance, $l$, to be measured (see
\cite{Adler},\cite{Baez},\cite{Requ1}). While the (technical) details
of the arguments given in the three papers are certainly slightly
different and vary with respect to the tightness and conclusiveness of
the steps in the respective lines of reasoning, the overall focus of
the papers is similar in spirit (see below).\\[0.3cm]
Remark: One should note that we were not aware of the two earlier
papers when we prepared our own contribution
\cite{Requ1}!\vspace{0.3cm}

As far as we can see, the original authors only reacted (in a quite
negative way) to the first paper \cite{Adler} (see \cite{Cam4} and
\cite{Ng5}). Apparently they considered the topic to be then settled
and, to our knowledge, did not even mention the later (and more
elaborate) accounts in \cite{Baez} and \cite{Requ1} (cf. e.g. the
recent \cite{Ng4}). Therefore we feel obliged to give a considerably
more careful and detailed account in the present work of our counter
arguments. This holds the more so as we feel the whole matter is of
extreme importance (both experimentally and theoretically).

Before we proceed with the more technical analysis, some general
remarks concerning the whole field and the logical structure of the
various arguments seem to be in order. Both the original analysis of
e.g. Ng et al. or Amelino-Camelia and our own contribution
(\cite{Requ1}) actually consist of roughly two parts (which are of
course related). For one, the Salecker-Wigner thought experiment, for
another, a semi-phenomenological theory of low-energy quantum gravity
or \tit{space-time foam}. We recently developed for example certain
aspects of such a theory in \cite{Requ2}, based on the
\tit{holographic principle}. The second half of \cite{Requ1} also
deals with this special topic and consequences drawn by e.g. Ng et
al. Furthermore, in \cite{Roy} we developed a theory of random metric
spaces and applied it to models of quantum space-time. We agree with
Amelino-Camelia that this is a very important and desirable enterprise
(cf. the abstract of \cite{Cam4}). It is however disputable and in
fact a different question if the claims concerning fundamental bounds
derived from the Salecker-Wigner thought experiment are really
correct. We think, our counter arguments have been too quickly brushed
under the carpet and that the situation is in fact considerably more
subtle. In the present paper we will only deal with this thought
experiment and its implications in order to keep the investigation
within reasonable size. We plan to treat the question of stochastic
fluctuations of space-time in a forthcoming paper.

We close this introduction with the mentioning of two particular
points which should be given a closer inspection as they are in fact
crucial for the logical coherence of the arguments being advanced in
favor of the various points of view. First, in these types of thought
experiments where extreme limit situations are studied (concerning the
very possibility of the experimental realisability), it is important
to check how large or small certain constants or parameters really
are, to what extent they can be freely chosen (e.g. in cases
where they are assumed to approach zero or infinity), or, on the other
hand, whether there exist practical or fundamental constraints. A
typical case in point is the habit to tacitly replace an estimate
containing the relational symbols $>$ or $<$ by $\gtrsim$ or
$\lesssim$ and then proceed by assuming without a more detailed
discussion that the upper or lower bound can actually be reached in
practice while a closer technical inspection would rather show that
this is not! possible and that the relation is more adequately
described by the symbol $\gg$ or $\ll$.

This problem becomes for example aparent if one replaces the only
approximatily correct continum models (e.g. elastic rods) describing
the devices, typically used in this context, by their more reliable
microscopic counterparts, based on the laws of many-body quantum
theory (cf. our sections 3 and 4). It is easy to make adhoc
assumptions about the possible physical parameters of these devices as
long as one does not go into their microscopic and quantum mechanical
details.

Second, and this concerns the second part of the usual argumentation,
that is, the relation of the Salecker-Wigner thought experiment to the
claimed fundamentality of the measurement bounds: It is clear that the
technical (thought) experiments alone are not sufficient to support
this claim. Both mentioned groups evidently seem to be intrigued by
the functional form of the terms $l^{1/2}$ or $l^{1/3}$, which
apparently remind them of versions of Brownian motion models (see
e.g. section 4 of \cite{Cam3}). As to the occurrence of these terms
one should say that the Salecker-Wigner experiment alone does by no
means suggest such a deeper connection to quantum gravity effects. It
is evident that these terms occur in certain expressions
because of the quantum-uncertainty induced movement of clock and/or
mirror. This is, in a sense, a quite trivial effect and does not seem
to have anything to do with Planck fluctuations as long as one does
not argue that quantum theory is a large scale consequence of quantum
gravity (which may in fact be the case).   

Be that as it may, the original authors argue that the fundamental
character of length fluctuations derives from the cooperative and
correlated behavior of the individual grains of space-time. The above
power laws would then suggest a rather mild form of correlation. In
the second part of our \cite{Requ1} we scrutinized these ideas and
came to a different conclusion, that is, the \tit{holographic
  principle} and other observations rather suggest an extremely strong
form of, what we called, \tit{anticorrelation} of the fluctuations of
the hypothetical individual grains of space-time. This means, the
individual fluctuations have the tendency of cooperating in such a way
that the total fluctuation in macroscopic regions remains small,
i.e. just the opposite of a central-limit or Brownian motion
behavior. We later discussed this peculiar fluctuation structure in
greater detail in \cite{Requ2} (see also \cite{Roy}).

That is, the real question in this context is the following, and this
goes beyond the question, mainly adressed in the present paper, which
primarily deals with the Salecker-Wigner thought experiment:
\begin{ob}Are the geometric fluctuations in the quantum vacuum near
  the Planck scale only weakly correlated, as suggested by the
  (Brownian-motion) results of Amelino-Camelia or Ng et al., or are
  they strongly anticorrelated as suggested by our own findings? The
  latter possibilty would entail that e.g. length fluctuations are
  essentially independent of the length to be measured. 
\end{ob}
\section{A Brief Review of the
  Salecker-Wigner-\\Amelino-Camelia-Ng-van Dam Thought Experiment}
In this section we will be very brief, only emphasizing certain more
relevant aspects, as the topic has meanwhile been described repeatedly
(apart from the original sources in e.g. \cite{Baez} or
\cite{Requ1}). We begin our analysis with a general remark.\\[0.3cm]
Remark: One should keep in mind that various of the more technical and
practical problems belonging to the special field of length and/or
time metrology, are not discussed in the cited papers and also not in
the following as this would become a quite cumbersome enterprise. To
this belongs for example the problem of the exact determination of the
arrival time of light pulses or individual photons, the inescapable
microscopical roughness of the surface of mirrors which neccessitates
the use of light with wave lengths which are sufficiently long so as
to average over this roughness of surfaces etc. In the following we
rather try to concentrate on the more fundamental problems. However,
we show that such a fundamental analysis may nevertheless lead to
interesting technical suggestions (see section 4.3).\vspace{0.3cm}

The original Salecker-Wigner thought
experiment deals (among other things) with the quantum-uncertainty of
length measurements in a gravitational field. As has been rightly
emphasized in \cite{Diosi}, its main concern was rather the
construction of tight nets of coordinate lines in general relativity
if quantum effects are included. Therefore their reliance on freely
falling clocks, mirrors etc. was quite reasonable, as this is natural
in this context.

Note that the definition of (true) spatial distance in e.g. a static
gravitational field is not completely trivial but nevertheless
straightforward (for a clear account see for example \cite{Landau},
the respective formulas can also be looked up in \cite{Requ1}). In the
Salecker-Wigner approach a light pulse is sent from a small freely
falling apparatus which also contains a clock towards an also freely
falling mirror where it is reflected. The distance can then be
inferred from the total arrival time needed, i.e.
\begin{equation}2l=t\cdot c        \end{equation}
with $t$ the elapsed time. 

If the quantum nature of clock and mirror is taken into account, their
positions at the respective arrival times of the light pulse are
uncertain by an amount
\begin{equation}\delta l+\delta v\cdot t= \delta l+\hbar/m\delta
  l\cdot l/c= \delta l+\hbar l/mc\delta l        \end{equation}
with 
\begin{equation}\delta l\cdot\delta p\geq (1/2)\hbar\quad , \quad
  \delta v=\delta p/m       \end{equation}
$m$ being the mass of clock or mirror, $\delta l$ the original
position uncertainty or, rather, the position uncertainty after the
emission of a light pulse; for convenience we discuss only the case of the clock. 
This yields a minimal uncertainty
\begin{equation}\delta l_{min}=(\hbar l/mc)^{1/2}=l_c^{1/2}\cdot l^{1/2}      \end{equation}
with $l_c$ the Compton wave length of clock or mirror.       

One can now try to minimize the uncertainty in length or position
measurement by making $m$ as large as possible. There are, obviously,
practical limits, for example if one wants to create a dense
coordinate net as e.g. in \cite{Wigner} or \cite{Diosi}. But there
exists also a fundamental limit given by the \tit{Schwarzschild-bound}
as has been exploited in \cite{Cam1} or \cite{Ng1}. One should note
that such Schwarzschild-type arguments were of course already used in
the past in related contexts. For one, the uncertainty in position
imparts also a fluctuation in the gravitational field and the metric
tensor (\cite{Cam1}). Furthermore, huge masses lead to a macroscopic
distortion of the gravitational field in the large. This, however,
represents in our view rather a correction and not! an uncertainty and
can be incorporated by a rigorous distance calculation as e.g.
described in \cite{Landau}. Really crucial seems to be, at first
glance, the Schwarzschild constraint.

If the geometric size of the clock-lightgun system (or mirror) is
given by $s$, a horizon will form around the clock (or mirror) if
\begin{equation}m\geq m_s:= const\cdot (c^2s/G)= const\cdot \hbar/c\cdot
  s/l_p^2 \end{equation} for some constant of order one. By inserting
this estimate in the preceding expression, Amelino-Camelia derives a
lower bound on the uncertainty, $\delta l$, of the form
\begin{equation}\delta l\geq const\cdot (l_p^2/s)^{1/2}\cdot l^{1/2}     \end{equation}
More specifically, one exploits the estimate in the opposite direction
in order that
the measurement device is able to function as expected. \\[0.3cm]
Remark: We will later comment on the relation of the size of the
clock, $s$, to the length, $l$, to be measured. Frequently the size is
assumed to be very small. This, however, does not seem to be necessary
in
our view.\\[0.3cm]
Jack Ng and van Dam (\cite{Ng1}) get a slightly different estimate by
using a \tit{light clock}. Again they make an estimate of the size of
the light clock which is in our view overly restrictive. In their
example, photons bounces back and forth in a cavity of size $b$. They
correctly argue that the smallest time interval one can resolve with
this clock is of order $t_b=b/c$. This induces a length uncertainty of
order $\delta l\gtrsim b$. They then however assume that the size of the
clock apparatus is also of size $b$, i.e. that the whole mass, $m$, of
the clock is squeezed into this (small) region of size $b$. We do not
see that this restrictive assumption is really necessary. We think,
one can envisage a massive but extended clock system containing e.g. a
small cavity of size $b$ while the size of the whole system, $s$, is
considerably larger (see below). They now conclude that it follows
\begin{equation}\delta l\geq b\geq l_s:= const.\cdot Gm/c^2       \end{equation}
i.e. with the rhs the corresponding Schwarzschild radius. They then
get
\begin{equation}\delta l\geq const\cdot l_p^{2/3}\cdot l^{1/3}         \end{equation}
Both estimates show the (at first glance) strange dependence of
$\delta l$ on the length, $l$, to be measured. But in our view this is
only the consequence of the particular experimental set-up of the
Salecker-Wigner experiment with its freely falling objects. It remains
to be shown that it is of a more fundamental significance.

While our technical analysis will start with the following section, we
want in this section to comment on a simple example Ng et al. give in
order to corroborate their strange result. Furthermore it supports our
suspicion that the possibility of strong anticorrelations in this
context is obviously not seriously taken into account by some authors
(whereas this possibility is sometimes mentioned in passing). The
reason is presumably that it seems to be not so easy to imagine
physical mechanisms which produce such strong effects; see however the
second part of \cite{Requ1} and the detailed analysis in \cite{Requ2}.

We begin with some, as we hope, clarifying remarks for readers not so
familiar with solid state physics. Both real and harmonic crystals,
which we discuss in more detail in the next section, are not stable in
one dimension. But as the periodicity in the harmonic model is put in
by hand via an explicitly given lattice constant, $a$, in the
1-dim. case, the unstable behavior is reflected by the divergence of
the fluctuations of the atomic positions around their equilibrium
values when the particle number, $N$, goes to infinity. That is, with
$u_i:=x_i-x_{i,0}$ we have in one dimension for non-vanishing
temperature:
\begin{equation}\langle u_i^2\rangle^{1/2}\sim
  N^{1/2}    \end{equation}
with $N$ the number of atoms in the chain. By the same token, with
$l=N\cdot a$ the average distance between, say, the atoms at positions
$x_{0,0}=0$ and $x_{N,0}=N\cdot a$, the respective distance fluctuation is
\begin{equation}\delta l\lesssim \langle u_0^2\rangle^{1/2}+ \langle u_N^2\rangle^{1/2}    \end{equation}
(apart from small possible boundary corrections which depend on the
boundary conditions being used).

In \cite{Ng1} Ng and van Dam argue that length fluctuations being
proportional to some simple fractional power of the length to be
measured are typical and natural and give the following example. They
mention a one-dimensional chain of $N$ ions connected by springs (an
example, they attribute to Wigner) in the high-temperature limit. With
$b$ the lattice constant and $\delta x_i:=x_i-x_{i-1}$ they argue that
\begin{equation}\langle (x_N-x_0)^2\rangle^{1/2}=\delta l\sim (N\cdot <\delta
  x_i^2>)^{1/2}=l^{1/2}\cdot (<\delta  x_i^2>/b)^{1/2}      \end{equation}
with $l=N\cdot b$ and $<\delta  x_i^2>$ being independent of the
position $i$ (modulo certain boundary conditions).

One should say this is a fairly unsurprising observation and does by
no means corroborate their general claim. For one, for very high
temperatures the individual atoms fluctuate almost independently
relative to each other. The above result is then nothing but the
well-known \tit{central limit theorem}. But even for ordinary
(non-vanishing) temperatures we get a similar result after some
calculations (see the next section). On the other hand, in the
following sections we mainly deal with the nature of \tit{zero-point
  motions} at temperature zero. For a one-dimensional harmonic crystal
at zero temperature we then get
\begin{equation}\delta l\sim (\ln N)^{1/2}     \end{equation}
On the other hand, we will see in the next section that in higher
dimensions atomic fluctuations remain small and finite and, by the
same token, fluctuations in distances. Therefore the one-dimensional
harmonic chain is rather exceptional and does not represent the
typical case.

The scenarios, described by the above cited authors rather prevail in
gases or other disordered or weakly correlated systems. We remind the
reader of the observation at the end of the introduction. In this
context we again want to mention our results in the second part of
\cite{Requ1} where we already discussed in some detail the harmonic
crystal and showed that it is exactly an example from the realm of
ordinary physics displaying these strong anticorrelations we mentioned
above. In this sense, we think, nothing really follows from this
example.

We conclude this section with some brief remarks concerning the
different scenarios employed by Amelino-Camelia, Ng et al., Adler et
al., Baez et al. or by ourselves, because this gives the motivation
for our detailed investigation into the behavior of real quantum
solids in this field of quantum gravity research. Most of the devices
employed are made of such stuff and we think, the at best
approximately correct continuum models do perhaps not! give the
correct results in these extreme (high-precision) situations.

The original authors essentially used these devices in the way
Salecker-Wigner used them, i.e. clocks, lightguns and mirrors are
designed and treated as freely falling, relatively small (but possibly
heavy) objects, the microscopic structure of which does not play a
crucial role and is more or less neglected. In \cite{Adler} the clock
is assumed to be somehow bound in a harmonic containing potential, the
physical nature of which is not openly indicated in detail. Its main
purpose is to keep the clock (or mirror) from wandering away under its
original momentum uncertainty.

Baez et al. give a more detailed account by assuming the clock being
fixed at the end of a (long) rod and then estimate its length
oscillations within the framework of continuum mechanics. In
\cite{Requ1} we fixed both clock/lightgun and mirror on a solid
understructure, the behavior of which we treated in a microscopic
quantum mechanical way. Now the relative mean-distance is fixed,
because neither clock nor mirror can wander away, but what is still
present (as in the other treatments) are the unavoidable quantum
fluctuations of (in our case) the individual atomes of the crystal
lattice. What is however now avoided to an at least large extent are
the strong constraints, resulting from the Schwarzschild bound, which
entered in the treatment of the original authors because they assumed
that the respective devices are quite small.

Most of the authors seem to follow the original idea of independently
located devices in space (i.e. clock, mirror, other objects). They
then automatically have to struggle with the quantum-uncertainty
induced movement of the objects which introduces these funny terms, we
have discussed above. On the other hand, Ng et al. and Amelino-Camelia
invoke the possible usefulness of gravity-wave interferometers. But as
far as we have understood the subject matter, these are very large and
massive complex devices with most of the equipment sitting on a rigid
extended under-construction. There exist of course certain parts which
are suspended or are able to oscillate, but nevertheless there
relative mean distances and positions are essentially fixed (see
e.g. \cite{Hough}). That is, in our view, these constructions seem to
resemble rather the experimental set-up we are suggesting, i.e. clock
and mirror being parts of a more or less rigid and complex measuring
device.

While we have the impression that for example Amelino-Camelia seems to
have the opinion that the presence of such extended bodies will
distort the gravitational field in a perhaps uncontrollable way, we
think, these perturbative effects can be incorporated in the
calculations. Anyhow, we have not found a really convincing argument
in favor of this pessimistic opinion. This holds the more so as
ultimately most of our equipment happens to be fixed to our planet
earth or to some other huge body and this is certainly the case for
the mentioned interferometers. Furthermore, after all we are in fact
interested in matters of principle and not in numerical details. That
means, the minute fluctuations of space-time are expected to occur not
only in free space but as well in solids and other equipment (remember
the solid cylinders of the first gravity-wave experiments).

In \cite{Requ1} we assumed clock (or mirror) to be fixed to the rigid
understructure by means of a trap (as in \cite{Adler}), which was
implemented by some oscillator potential. We meanwhile think this
additional source of uncertainty is not really necessary. In the
following we prefer to regard the clock (and mirror) as being
integrated parts of the rigid body itself. The task is then to
carefully check the various occurring physical parameters of the
different models, employed in this context, as to the possibilty of
choosing them in such a way so that the final length fluctuations we
are interested in become as small as possible.

In this respect two, at least in our view, different questions have to
be dealt with. First, the claim of Amelino-Camelia and Ng et al. that
length fluctuations really have an intrinsic and unavoidable
dependence on the length to be measured, i.e.
\begin{equation}\delta l\gtrsim l^{1/2}\quad\text{or}\quad \delta l\gtrsim
  l^{1/3}    \end{equation}
for length scales which can in principle be experimentally
observed. Second, good quantitative estimates of the numerical degree
of length uncertainty in the different experimental setups. We
emphasize this latter point as we suspect that the existing estimates
are not very reliable and that certain assumptions are perhaps too optimistic.
\section{Solid State Physics meets Quantum Gravity}
In the various thought experiments which have been introduced in the
field we are discussing, equipment has been employed which is in the
last analysis of the nature of quantum many-body systems. As we are, a
fortiori, employing this measurement equipment in very extreme
situations, we think it is reasonable to take its microscopic
many-body nature really into account and not simply regard the
measuring devices as being essentially classical or structureless
objects.

Take for example the one-dimensional rod introduced in \cite{Baez},
the behavior of which is discussed, at least in the first steps,
within the framework of classical continuum mechanics. Its length is
denoted by $x$, the velocity of sound by $c_s$, the elastic modulus by
$Y$, its mass by $m$. With $\rho=x/m$ the mass density in one
dimension, the general formula for $c_s$ is
\begin{equation}c_s=(Y/\rho)^{1/2}=(Y\cdot x/m)^{1/2}       \end{equation}

With the apriori bound ($c$ being the velocity of light)
\begin{equation}c_s\leq c      \end{equation}
and the heuristic association of the rod with a harmonic oscillator
having spring constant
\begin{equation}k=2Y/x \end{equation}
 Baez et al. finally get the
formula for the zero point length fluctuation of the rod via the
associated harmonic oscillator model.
\begin{equation}\Delta x\gtrsim (\hbar x/mc)^{1/2}    \end{equation}
with, at first glance, an explicit dependence of $\Delta x$ on $x$
while written in the form 
\begin{equation}\Delta x\gtrsim (\hbar/\rho c)^{1/2}     \end{equation}
one sees that the length fluctuation, calculated in some oscillator
ground mode is actually independent of the length of the rod (at least
as long as the \tit{Schwarzschild constraint} has not been
introduced).\\[0.3cm]
Remark: We surmise that the authors used some form of Hooke's law in
this derivation. One should note that the correct form of Hooke's law
reads
\begin{equation}\Delta l/l=Y^{-1}\cdot F/A= Y^{-1}\cdot\sigma      \end{equation}
with $F$ the applied force, $A$ the area of the cross section and
$\sigma$ the tension in the rod. The other variant one frequently
finds in the literature
\begin{equation}\Delta l=k^{-1}\cdot F    \end{equation}
has the disadvantage of hiding the explicit dependence of $k$ on the
length of the rod, i.e. we have
\begin{equation}k=Y/l\cdot A   \end{equation}

We will show in the following that this (heuristic) fluctuation
result, based to a large part on classical physics, coincides with the
rigorous microscopic result (for zero temperature!) apart from an
(inessential) factor of the form $(\ln N)^{1/2}$ with $N$ the number
of atoms. Note again that in a strict sense the one-dimensional
harmonic crystal is not stable in the limit $N\to\infty$. That means,
the fluctuations of the individual atoms diverge in this limit (but in
an extremely slow manner in the case $T=0$). We will come back to the
approach of Baez et al. in the following section in connection with
the Schwarzschild-constraint and the \tit{Hoop-conjecture}. We will
then see more clearly the possible weaknesses of such continuum
models.

As a typical candidate for a true many-body system serving as a model
for the possible macroscopic measuring devices we will employ in the
following we now discuss the harmonic crystal and the position
fluctuations of its atoms. We will later see that it may be
advantageous to also use equipment which is not entirely made of this
rigid crystallic structure but contains also parts which are capable
of absorbing and damping various sources of external and internal
noise. To begin with, we assume the crystal to be cooled down to
$T=0$. The qualitatively different behavior for non-vanishing $T$ will
be discussed at the end of this section. This means, that only the
so-called \tit{zero-point motion} of the atoms is taken into
account. Various aspects of this model are e.g. discussed in
\cite{Ziman} or \cite{Mermin}, see also \cite{Becker}, the classical
source being \cite{Peierls}. But as the results are frequently widely
scattered in these books and as the quantum fluctuation results we are
interested in, are either not explicitly given or difficult to find,
we will provide the necessary formulas in the following. One should
furthermore note that we are sometimes cavalier concerning (in this
context) uninteresting prefactors of order one.

In the following discussion we assume that the crystal as a whole is
fixed in the respective reference system or, rather, we consider it
relative to its center-of-mass system. Put differently, we neglect,
for the time being, the purely translatory mode. So, let $\mbf{u}_i$
be the momentary elongation of the $i$-th atom from its equilibrium
position and $N$ the number of constituent atoms. We then have in the
quantum case
\begin{equation}N^{-1}\cdot\sum_{i=1}^N
  <\mbf{u}_i^2>=N^{-1}\cdot\sum_{\mbf{k},s}
  \hbar/(2M_0\,\omega_{\mbf{k},s}) \end{equation} with the brackets in
our present context ($T=0$) denoting quantum averages. For
non-vanishing temperatures the formula has a slightly different
form. The sum on the rhs runs over the first \tit{Brillouin zone} of
the crystal and over the different possible \tit{phonon branches}
$\omega_{\mbf{k},s}$. $M_0$ is the mass of the lattice atoms. For
$\mbf{k}\to 0$ we have
\begin{equation} \omega_{\mbf{k},s}\sim c_s(\mbf{k}/|\mbf{k}|)\cdot\mbf{k}      \end{equation}
with $c_s$ the (in general) branch and direction dependent velocity of sound.

To evaluate the sum we make some (harmless) approximations. We
restrict ourselves to a single phonon branch, extend the linear
dispersion law of the (accoustic) phonon branch up to the boundary of
the Brillouin zone, assume $c_s$ to be independent of the direction
and replace the first Brillouin zone by the so-called Debeye
sphere. Furthermore, in space dimension greater than one we replace
the sum by an integral, using the conversion factor (with $a$ some
lattice constant)
\begin{equation}\sum_{\mbf{k}}=Na^3/(2\pi)^3\int         \end{equation}
We finally get in three or two dimensions
\begin{ob}For the fluctuation of the position of the lattice atoms in
  a harmonic crystal at $T=0$, i.e. only zero-point fluctuations being
  taken into account, we get
\begin{enumerate}
\item 2,3-dim.:
\begin{equation}\Delta u_i\approx const\cdot (\hbar a/c_sM_0)^{1/2}       \end{equation}
\item 1-dim.:
\begin{equation}\Delta u_i\approx const\cdot (\hbar
  a/c_sM_0)^{1/2}\cdot (\ln N)^{1/2}      \end{equation}
\end{enumerate}
with $const$ being of order one. This implies that the fluctuations of
the distance between two arbitrary lattice sites is of the same order.
\end{ob}
Proof: The three and two dimensional case follows from our above
formulas. The one-dimensional case has to be treated slightly
differently. It is more appropriate to directly calculate the discrete
sum 
\begin{equation}(\hbar/M_0)\cdot N^{-1}\sum_{Brill.zone}(c_s\cdot |\mbf{k}|)^{-1}     \end{equation}
as the integral version is singular at $\mbf{k}=0$. This yields
\begin{equation}\Delta u_i^2\approx  (\hbar a/c_sM_0)\cdot
  N^{-1}\cdot\sum_{l=1}^N (2\pi l/Na))^{-1}\approx const\cdot (\hbar
  a/c_sM_0)\cdot\ln N     \end{equation}
Remark: In \cite{Requ1} we got already similar results (for space
dimension greater than one) with the help
of a slightly different reasoning. Our main goal in \cite{Requ1} was
however to show how natural strong anticorrelations among individual
position fluctuations are already in ordinary physics and that the
standard Brownian-motion type results or variants thereof, which all
are somehow inspired by the central limit behavior, and which are
typically invoked in this context, cannot always be expected.

It is now important to investigate the range within which the
occurring physical parameters can be chosen and, furthermore, if they
can be independently chosen. Note in this context that in the harmonic
crystal model the lattice constant is put in by hand. In the true
many-body situation the periodicity of certain states has in principle
to be calculated (which is quite difficult). Furthermore, the lattice
constant is expected to change if $M_0$ or e.g. the temperature is
varied.

In a first step one can try to make $c_s$ as large as possible. We
evidently have an apriori upper bound 
\begin{equation}c_s\leq c \end{equation} with $c$ the velocity of
light, which is however quite crude as typical values for $c_s$ are of
order $10^3[m]/[s]$. As to the lattice constant $a$, it seems to be
difficult in ordinary matter to have it smaller than the average
distance in dense nuclear matter (e.g. neutron stars), i.e. we may
assume
\begin{equation}a\geq a_0\approx 10^{-15}[m] \end{equation} There
exists another relation between $c_s,M_0,a$ and the coupling constant,
$\alpha$, of the harmonic oscillator potential between neighboring
atoms
\begin{equation}c_s=(\alpha/M_0)^{1/2}\cdot a      \end{equation}
which, when choosing the extreme values $c_s=c$ and $a=a_0$, yields a
relation between $\alpha$ and $M_0$.

We see that the following estimates hold.
\begin{ob}In ordinary matter the expression $(\hbar\,
  a/M_0\,c_s)^{1/2}$ is lower bounded by
\begin{equation}(\hbar\,
  a/M_0\,c_s)^{1/2}\gtrsim l_c(M_0)^{1/2}\,a^{1/2}\gtrsim l_c(M_0)^{1/2}\,a_0^{1/2}     \end{equation}
with $l_c(M_0)$ the Compton wavelength of the lattice atoms.
\end{ob}
It is instructive to calculate this bound numerically. With
$M_0\approx 10^{-25}[kg]$ for ordinary atoms, we get
\begin{equation}l_c(M_0)^{1/2}\cdot a_0^{1/2}\approx 10^{-16}[m]     \end{equation}
On the other hand, for ordinary matter with $a\approx 10^{-10}[m]$ and
$c_s\approx 10^3[m]/[s]$ we get
\begin{equation}(\hbar\,
  a/M_0\,c_s)^{1/2}\approx 10^{-11}[m]     \end{equation}
\begin{conclusion}A reasonable lower bound for the length fluctuations
  in three, two, or one space dimensions is (with $(\ln N)^{1/2}=O(1)$)
\begin{equation}\Delta u_i\gtrsim 10^{-16}[m]     \end{equation}
\end{conclusion}

This lower bound on $\Delta u_i$ is obviously still far above the
Planck scale, but it is a reliable value as long as we do not take
special measures. One should compare this bound with the bound of Baez
et al. for the one-dimensional elastic rod. One should note that in
\cite{Baez} only a single ground frequency was used in the idealized
model. In our microscopic rigorous approach we integrated over all
occurring phonon frequencies. The effect is however only an extra
numerical prefactor.
\begin{ob}Relating the estimate in the approach of Baez et al. with
  our rigorous microscopic calculation we get (without the additional
  Schwarzschild constraint)
\begin{equation}\Delta x\gtrsim (\hbar\, x/m\,c)^{1/2}\gtrsim
  (\hbar\,N\cdot a_0/N\cdot M_0\,c)^{1/2}= l_c(M_0)^{1/2}\cdot a_0^{1/2}     \end{equation}
i.e., the two estimates give roughly the same value. It is however
crucial that the rhs of the above estimate shows that the length
fluctuation is completely independent of the parameters $x$ and/or
$m$. So there seems to be no room left to make $\Delta x$ small by
choosing $x$ or $m$ appropriately (see the next section).  
\end{ob}

Before we proceed we mention the corresponding results for
non-vanishing temperature. In this case we have 
\begin{equation}N^{-1}\cdot\sum_i <u_i^2>=N^{-1}\cdot\sum_k
  (\hbar/2M_0\,\omega_k)\,\cot (\beta\,\hbar\,\omega_k/2)      \end{equation}
with $\beta$ the inverse temperature. For small $\mbf{k}$ we get
\begin{equation}\cot (\beta\,\hbar\,\omega_k/2)\approx
  2/\beta\,\hbar\,c_s\cdot |\mbf{k}|^{-1}     \end{equation}
and
\begin{equation}N^{-1}\cdot\sum_i
  <u_i^2>=N^{-1}\cdot\sum_k(\beta\,M_0\,c_s)^{-1}\cdot
  |\mbf{k}|^{-2}  \end{equation}
\begin{conclusion}For $T\neq 0$ we get the estimate
\begin{equation}\Delta u_i\approx  (\beta\,M_0\,c_s)^{-1}\cdot a    \end{equation}
in three or two dimensions and
\begin{equation}\Delta u_i\approx   (\beta\,M_0\,c_s)^{-1}\cdot
  a\cdot N^{1/2}    \end{equation}
in one dimension
\end{conclusion}

To sum up what we have learned in this section; we have seen that, due
to the atomic structure of ordinary matter, it is rather academic to
make incompatible assumptions in certain thought experiments as to
various occurring physical parameters of objects or equipment to be
employed in some of the arguments. It is for example problematic to
assume that very small but sufficiently heavy objects do actually
exist. It may turn out that in the far future some exotic matter may
be found having such properties but at the moment it seems to be
difficult to pack ordinary matter denser than with interatomic
distance $a_0\approx 10^{-15}[m]$. If we assume that a typical atomic
mass is of order $M_0\approx 10^{-25}[kg]$, we have the following
constraint.
\begin{ob}For the size, $s$, and mass, $m$, of a typical object in our
  discussion we have the following relation (with $N$ the number of
  atomic constituents):
\begin{enumerate}
\item If the object is essentially three-dimensional we have
\begin{equation}s\gtrsim N^{1/3}\, a_0\quad , \quad m= N\cdot M_0      \end{equation}
or
\begin{equation}s\gtrsim (m/M_0)^{1/3}\, a_0       \end{equation}
\item In two dimensions we get
\begin{equation}s\gtrsim (m/M_0)^{1/2}\, a_0     \end{equation}
\item In one dimension for a rod-like shape we get
\begin{equation}s\gtrsim N\, a_0\quad , \quad m= N\cdot M_0      \end{equation}
and
\begin{equation}s\gtrsim (m/M_0)\, a_0       \end{equation}
\end{enumerate}
\end{ob}
These bounds will have certain consequences for the discussion in the
following section. 
\section{Commentary on the various Thought Experiments}
In the light of our previous observations we will comment on the
thought experiments by Baez et al. and Ng et al and will compare them
with our own approach.
\subsection{The Modified Thought Experiment of Baez et al. and the
  Hoop Conjecture}
In \cite{Baez} the Schwarzschild constraint is used for an essentially
one-dimensional rod of length $x$ by invoking the so-called \tit{Hoop
  conjecture} (see e.g. \cite{Thorne} or \cite{Wheeler}). If $m$ is the
mass of the rod it roughly says that a horizon will form around the
rod if
\begin{equation}m\geq m_s= const\cdot c^2\,x/G     \end{equation}
with some constant of order one.

To make the length fluctuation of the rod as small as possible Baez et
al. chose the length of the rod to be roughly the size of the
corresponding Schwarzschild radius or, rather, a little bit larger,
that is
\begin{equation}x\approx l_s=const\, m\,G/c^2    \end{equation}
Then the product $x^{1/2}\,l_c^{1/2}$, occurring in their derivation,
would become approximately
\begin{equation}l_s^{1/2}\,l_c^{1/2}=l_p \end{equation} and they
finally concluded (by incorporating the additional uncertainty of the
center of mass of the rod and by choosing $m$ arbitrarily large)
\begin{equation}\Delta x \gtrsim l_p       \end{equation}

We already derived in the preceding section a lower bound on $\Delta
x$ for the particular experimental set-up used by Baez et al and which
is completely independent of $x$ and/or $m$ but is much larger than the
Planck length, that is
\begin{equation}\Delta x \gtrsim l_c(M_0)^{1/2}\cdot a_0^{1/2}      \end{equation}
This implies, that something must be wrong in the reasoning of Baez et
al. We will show now that it is not possible for the length of an
essentially one-dimensional rod made from ordinary matter to come near
the Scharzschild radius of the rod. At the end of the preceding
section we got a relation between mass and length of a one-dimensional
rod
\begin{equation}N\cdot a=l\gtrsim m/M_0\cdot a_0= N\cdot
  a_0 \end{equation}
with $a$ the real lattice constant, $a_0$ its minimal value. This implies
\begin{equation}l/l_s=a\,c^2/(M_0\,G) \gtrsim a_0\,c^2/(M_0\,G)     \end{equation}
With our standard assumptions $M_0\approx 10^{-25}[kg],a_0\approx
10^{-15}[m]$ we get
\begin{equation}M_0\,G/c^2=l_s(M_0)\approx 10^{-53}[m]\quad\text{and hence}\quad
  l/l_s\gtrsim 10^{38}    \end{equation}
\begin{ob}For ordinary matter the linear extension of e.g. a rod
  exceeds its Schwarzschild radius by a factor of $\gtrsim 10^{38}$.
  One should note that this holds for the rather extreme parameter,
  $a_0$, we have chosen. For a more realistic parameter the factor
  happens to be even larger.
\end{ob}
\begin{conclusion}For the above term $x^{1/2}\cdot l_c^{1/2}$ it holds
\begin{equation}\Delta x\gtrsim x^{1/2}\cdot l_c^{1/2}\gtrsim
  10^{19}l_s^{1/2}\cdot l_c^{1/2}=10^{19}\,l_p   \end{equation}
That is, there seems to be no chance that the length fluctuations of a
one-dimensional rod really come near the respective Planck value.
\end{conclusion}
This is a case in point for what we said in the introduction about
estimates using frequently the symbol $\gtrsim$ where rather the
symbol $\gg$ would be appropriate.

On the other hand, for three or two dimensions we got lower bounds at
the end of the preceding section of the kind ($s$ being the linear
extension of the object)
\begin{equation}s\gtrsim (m/M_0)^{1/3}\cdot a_0\quad , \quad  s\gtrsim (m/M_0)^{1/2}\cdot a_0   \end{equation}
yielding
\begin{equation}s/l_s\gtrsim const\, (c^2/M_0^{1/3}\,G)\cdot a_0\cdot
m^{-2/3}\quad , \quad s/l_s\gtrsim const\, (c^2/M_0^{1/2}\,G)\cdot a_0\cdot
m^{-1/2}\   \end{equation}
respectively. Setting $s/l_s=1$ on the lhs yields:
\begin{conclusion}While it seems to be impossible to confine a
  rod-like object of ordinary (atomic) matter within its Schwarzschild
  sphere or to come at least near this goal, this can be achieved in
  two or three space dimensions for sufficiently large mass. For three
  dimensions the mass scale such that $s=l_s$ is $m=m_s\approx
  10^{30}[kg]$ which is approximately the mass of the sun. The
  Schwarzschild radius is $s\approx l_s\approx 10^3[km]$. This result
  holds for the assumed extreme limit value $a_0$ we have chosen as a
  lower bound. Note that for a fixed value of the parameter $a$ both
  $l_s$ and $m_s$ are also fixed by the above formulas. Furthermore,
  for an $a>a_0$ both $m_s$ and $l_s$ become also larger.
\end{conclusion}
Remark: Note that in general, by neglecting the atomic microscopic
structure of matter, we can calculate for each given $s$ the
Schwarzschild mass, $m_s$, so that $s=l_s$. In the above calculations
we assumed that the average density or, put differently, the average
interatomic distance, $a$, is fixed or even
minimal. i.e. $a=a_0$. Then we get another relation between size and
mass and the identity $s=l_s$ can only hold for a single mass value.

\subsection{A Comparism of the Ng-van Dam Thought Experiment with our
  Approach}
The approach of Ng et al. (see section 2) is based on the
Salecker-Wigner method of freely falling (small) clocks and
mirrors. In a first step one gets
\begin{equation}\delta l\gtrsim l^{1/2}\cdot l_c^{1/2}          \end{equation}
In a second step they add the certainly correct assumption that the
size, $s$, of the clock has to be larger than its own Schwarzschild
radius. They then make however an additional assumption which in our
view is too restrictive. For the time measurement they choose a
so-called \tit{light-clock} in which a photon bounces between the
mirrored walls of a cavity. This is also reasonable as $c$ represents
a limiting velocity and a large velocity makes the period of the clock
short which, by the same token, is responsible for a technical lower
bound on the uncertainty of length measurement, i.e. we have
\begin{equation}\delta l_{tech.}\gtrsim b          \end{equation}
with $b$ the diameter of the cavity. They then however make the
assumption that the size of the clock is roughly of the order of the
diameter of the cavity, $b$. They hence get
\begin{equation}\delta l\gtrsim b\approx s\gtrsim l_s= const\cdot G\,m/c^2         \end{equation}
with $m$ the mass of the clock (or mirror). That is, they assume that the
whole mass of the clock is concentrated within a sphere roughly of the
size $b$. Combining the two estimates they finally get
\begin{equation}\delta l\gtrsim l^{1/3}\,l_p^{2/3}       \end{equation}

This may be contrasted with the estimate by Amelino-Camelia (see
section 2) in which no light clock was explicitly used:
\begin{equation}\delta l\geq const\cdot l_p\cdot (l/s)^{1/2}   \end{equation}
and in which the size of the clock-lightgun system still explicitly
appears. The form of the Ng-van-Dam estimate seems to convey a deep
(functional) relation between length measurement and Planck scale
physics, in particular as it contains the symbol $\gtrsim$ instead of,
say, $\gg$. We first should investigate if this connection does really
exist or, on the other hand, if it is only apparent.

We have seen that, first of all, in the approach of Ng et al. the
precision of length measurement is fundamentally limited by the period
of the light clock or, by the same token, by $b$. The smallest
conceivable clock of the kind Ng et al. are envisioning is, in our
opinion, a clock consisting of one atom or atomic nucleous. With our
rough approximation, $a_0$, we thus get
\begin{equation}\delta l\gtrsim \delta l_{techn.}\approx b\gtrsim a_0=10^{-15}[m]    \end{equation}
We previously calculated $l_s$ for such an atom and got
\begin{equation}l_s(M_0)\approx 10^{-53}[m]      \end{equation}
That is, for such an atomic clock its natural size exceeds its
Schwarzschild radius by many orders of magnitude, put differently, for
such a clock it would be inappropriate to use the symbol $\gtrsim$ in the
respective estimates. 

On the other hand, we derived in the preceding sections a relation
between the size of an object, made from ordinary matter, and its
mass. In three dimensions it reads (by assuming $a=a_0$):
\begin{equation}s\gtrsim (m/M_0)^{1/3}\cdot a_0 \end{equation} That
is, if we make the clock heavier, the Schwarzschild radius would also
increase but, by the same token, the size of the clock would increase
too. As $l_s$ grows linearly with $m$ while $s$ is proportional to
$m^{1/3}$ in three dimensions, there exists a unique value where $l_s$
and $s$ become identical. In the Ng-van Dam approach the size of the
clock is however rigidly related to the parameter $b$ which is a lower
bound to the uncertainty $\delta l_{techn.}$ which is smaller than
$\delta l$. So we arrive at a certain dilemma as the uncertainty in
time measurement would also increase with mass and size of the light
clock.
\begin{conclusion}In our view it is reasonable to use clocks with the
  parameters $s$ and $b$ being decoupled, that is, one should use
  clocks with the diameter of the mirrored cavity as small as possible
  in order to make $\delta l_{techn.}$ as small as possible but making
  their mass and, by the same token, their size large. Or, what seems
  to be even better, to use clocks with $l_{techn.}$ not limited by
  som geometric parameter $b$.
\end{conclusion}
Remark: As we are no expert in time metrology we do not know if there
perhaps exist ingenious methods to make the period of the time clock
shorter than the value we assumed, i.e. $\Delta t\approx
10^{-15}[m]/c$. This seems to be, at least in our view, difficult for
the type of light clock Ng et al. are employing but may be possible
for other types of clocks. One can learn from e.g. the analysis in
\cite{Hough} that the sensitivity of the modern interferometers can be
increased by various ingenious methods, but this seems to apply rather
to the observation of the (qualitative) change in interference
patterns, not so much to the exact measurement of distances.

 \vspace{0.3cm}

If we loosen the connection between the size of the mirrored cavity,
$b$, and the size of the whole clock system, $s$, we have more
possibilities. We can try to make $b$ as small as we can, or even
better, use a different sort of clock, and, on the other hand, make
$s$ as large as possible in order to avoid the Schwarzschild
constraint while we make $l_c$ as small as possible. We then fall back
on the relations derived in section 2, i.e.
\begin{equation}\delta l\gtrsim l_c^{1/2}\,
  l^{1/2}\quad ,\quad m\leq const\,(c^2\,s/G)=m_s     \end{equation}
as long as we insist on independent freely falling clock and mirror
systems. One should however remark that so far these devices are
treated as essentially structureless objects. The kind of internal
fluctuations being always present in these objects if their quantum
nature is taken into account has been treated in section 3 and will
further be treated in the following.

Inserting now (as e.g. Amelino-Camelia did) $m\approx m_s$ in
$l_c=\hbar/m\,c$, we get
\begin{equation}\delta l\gtrsim const\cdot l_p\,(l/s)^{1/2}     \end{equation}
From our previous calculations with $a=a_0$ we learned that both $m_s$
and $s$ are of considerable size. On the other hand, we think, this is
not totally unrealistic as in our framework these devices can be
considered to be more or less rigidly fixed onto for example the earth
itself. This is certainly the case, as we already emphasized above,
for the large interferometers the authors themselves invoked in their
arguments. That is, there is in our view no real need to resort to
small clocks and mirrors. Even if the length to be measured is small,
the clock-mirror system can ultimately be taken to have the size of
the earth.
\begin{conclusion}With clock and mirror being parts of some large
  devices which, on their side, being rigidly attached to e.g. the
  earth itself, both the Schwarzschild-constraint and the
  wandering-away effect can be essentially avoided so that, in
  the end, we get at least thought-experimentally an estimate of the
  kind
\begin{equation}\delta l\gtrsim const\cdot l_p    \end{equation}
Even if in practice the term $const$ may not really be of order one
but some small power of ten, the result is certainly independent of
the length $l$ itself.
\end{conclusion}
\subsection{The Statistical Mechanics of Relative Position
  Fluctuations of the Components of Large Measuring Devices}
What we and also the other authors have so far only superficially
discussed are the relative position fluctuations of parts of a larger
device relative to each other or relative to the larger device they
are embedded in. If, for example, the mirror is part of a larger
device or is used as a component in a clock system, the uncertainty of
distance mesurement is of course enhanced by the unavoidable
statistical movements of these parts relative to each other. To
formulate this problem in a more general way we will analyse in the
following the relative statistical movement of parts of a larger
many-body system, more precisely, of the respective centers of mass or
of the movement of the center of mass of a subsysten with respect to
the total system.

One should note that there exist in the literature various quite
heuristic statements concerning this point which are however not
satisfying in our context as they usually only apply to freely moving
objects. In our context the subsystem is in contact with a larger
system and, furthermore, there exist delicate and even long-range
correlations among the constituents of the object. Under such
conditions the problem is no longer totally trivial.

In section 3 we got already estimates on the individual fluctuations
of the atoms of a crystallic body. At zero temperature and ordinary
densities and velocities of sound we had roughly
\begin{equation}\Delta\,u_i\approx 10^{-11}[m]     \end{equation}
while for the extreme values, $c_s=c\quad a=a_0$, we got
\begin{equation}\Delta\,u_i\approx 10^{-16}[m]           \end{equation}
In \cite{Requ1} we already introduced the idea to attach e.g. clock
and/or mirror to larger parts of the under-structure in order to
further reduce the degree of position fluctuations, as in general the
center of mass of a subsystem, containing itself a substantial number
of atoms, is expected to display a smaller degree of fluctuations than
its individual constituent atoms. The quantitative analysis will
however depend on the general context.

Let us start with our standard example, the harmonic crystal. We will
see that in this case the problem turns out to be quite intricate. We
take a subcluster, $S$, of, say, $N$ atoms in the crystal with $N\gg
1$. The corresponding center of mass coordinate is
\begin{equation}\mbf{R}=\sum_{i=1}^N\,M_0\,\mbf{x}_i/N M_0          \end{equation}
The expected fluctuation of this coordinate can then be written as
\begin{equation}\langle (\mbf{R}-\mbf{R}_0)^2\rangle=N^{-2}\,\langle(\sum_{i=1}^N(\mbf{x}_i-\mbf{x}_{i,0}))^2\rangle         \end{equation}
with $\mbf{x}_{i,0}$ the equilibrium positions of the atoms and
$\mbf{R}_0$ the corresponding position of the center of mass. This
yields
\begin{multline}\langle
  (\mbf{R}-\mbf{R}_0)^2\rangle=N^{-1}\,\langle\,N^{-1}\,(\sum_{i=1}^N\mbf{u}_i)^2\rangle=\\
  N^{-1}\,(N^{-1}\sum_{i=1}^N\mbf{u}_i^2+2N^{-1}\,\sum_{i\neq
    j=1}^N\,\mbf{u}_i\cdot \mbf{u}_j)
      \end{multline}

As
\begin{equation}N^{-1}\,(\sum_{i=1}^N\mbf{u}_i)^2)=\Delta
  u_i^2\approx const\,(\hbar\,a/c_s\,M_0)       \end{equation}
(see section 3), the first term would essentially yield the result
which also follows from general handwaving arguments, i.e.
\begin{equation}\Delta\mbf{R}\approx N^{-1/2}\, \Delta \mbf{u}_i      \end{equation}
Problematical is however the second sum, as we know that the
$\mbf{u}_i$ happen to be long-range correlated in a crystal with the
correlations in 3-dim. only decaying in leading order proportional to
$|\mbf{x}_i-\mbf{x}_j|^{-1}$. On the other hand, we know that there is
a tendency of an oscillating behavior of correlations, that is, to
some extent the individual terms may compensate each other. But it
is very difficult to estimate this in a rigorous way.

The whole section 4 of \cite{Requ1} was devoted to this point in
connection with the question of (anti)correlations in the geometric
fluctuations of space-time on the Planck scale. We furthermore
mentioned in that section some older literature where such questions
have been systematically treated in the framework of statistical
mechanics and quantum field theory. The general problem consists in
estimating the behavior of autocorrelations of certain space-integrals
(or sum) over some physically relevant (operator)density, $q(x)$, i.e.
\begin{equation}Q_V:=\int_V\,q(x)d^3x\quad\text{or}\quad \sum_V\,q(x_i)      \end{equation}
In many cases one is interested in the behavior of $\langle Q_V\cdot
Q_V\rangle$ when $V$ becomes large or approaches the whole space,
$\R^3$. Obviously the correlation function
\begin{equation}\langle q(x)\,q(y)\rangle\quad\text{or rather}\quad
  \langle (q(x)-\langle q\rangle)\,(q(y) -\langle q\rangle)\rangle    \end{equation}
enters in this expression with $\langle q\rangle$, the expectation of
$q(x)$, being subtracted.

If the individual fluctuations are uncorrelated or only weakly
correlated (an integrable decay of correlations is sufficient) we get
a behavior
\begin{equation}\langle Q_V\cdot Q_V\rangle \sim R^3     \end{equation}
in 3-dim. with $R$ the diameter of the integration volume. The
situation frequently becomes better if certain covariance properties
are present (for example, $q(x)$ being the zero-component of a
conserved current); see section 4 of \cite{Requ1}. On the other hand,
if the correlations are of long-range character, the situation becomes
more complicated. Such a problem was for example analysed in
\cite{Requ3} section 3 or \cite{Requ4} section 5 in the field of
the statistical mechanics of phase transitions.

In that case we got roughly a result that
\begin{equation}\langle Q_V/V\cdot Q_V/V\rangle\lesssim R^{-1}      \end{equation}
for correlations decaying weakly like
\begin{equation}\langle q(x)\,q(y)\rangle\sim |x-y|^{-1}    \end{equation}
in 3-dim. and with $q(x)$ normalized to $\langle q(x)\rangle=0$. In
the above fluctuation result possible anticorrelation effects
(i.e. oscillations) are not included, only the decay property has been
used. That is, it is possible that the situation is actually better but
we do not know for sure.

Replacing now $q(x)$ by our $\mbf{u}_i$ and the integral by the
corresponding sum, this result can be taken over for the calculation
of the fluctuation of our center of mass variable, i.e. we have
\begin{conclusion}As the atomic position fluctuations in our crystal
  are long-range correlated, we get the rigorous bound
\begin{equation}\langle (\mbf{R}-\mbf{R}_0)^2\rangle^{1/2}\lesssim
  N^{-1/6}\cdot \Delta u_i   \end{equation}
with $\mbf{R}$ denoting the center of mass of some macroscopic part of
the whole crystal and $R\sim N^{1/3}$.
It is however possible that the estimate is better if
(anti)correlations do effectively cooperate.
\end{conclusion}

We learned however from our analysis in section 4 of \cite{Requ1} that
one can reduce the noise of the position fluctuations in solids, that
is, by the same token, in our measuring devices, if one uses
substances with short-range position fluctuations. Macroscopically
these short-range correlations work as damping mechanisms. So our idea
is to embed e.g. clocks and mirrors in components of the total
measuring system which display short-range correlations. These may be
for example viscous fluids or some dissordered systems. For such
systems we can use the results in section 4 of \cite{Requ1}. 
\begin{conclusion}For systems with e.g. integrable correlations we get
  the much better result for the center of mass motion
\begin{equation}\langle (\mbf{R}-\mbf{R}_0)^2\rangle^{1/2}\sim
  N^{-1/2}\cdot \Delta u_i      \end{equation}
with $\Delta u_i$ being of atomic order. This implies that for
sufficiently large $N$ the random movement of mirrors and clocks
become very small compared to typical atomic values and may come near the
Planck scale under ideal conditions.
\end{conclusion}

\section{Commentary}
To sum up what we have finally attained; we have shown that by using
compound systems as measuring devices, in which critical parts like
clocks and mirrors are embedded in components which effectively damp
the unavoidable random fluctuations of atomic positions via
short-range correlations, we can reach, at least in principle, a level
of precision regarding distance measurements, which may come near the
Planck level. In any case, we think, we have convincingly shown that
experimentally there is no indication that the precision of distance
measurements displays a functional dependence on the distance to be
measured, as has been claimed by e.g. Amelino-Camelia and Ng-van Dam.

These authors attributed this dependence to some Brownian-motion like
behavior of the geometric fluctuations in the micro-structure of
space-time. This is certainly a very interesting topic, but we will
show in a forthcoming paper (and have already argued in this direction
in previous work, cited above) that due to strong anticorrelations
these microscopic fluctuations have rather the tendency to compensate
each other, so that in the end we get a result which corroborates our
above analysis.\\[0.5cm]
{\small Acknowledgement: Fruitful discussions with H.J. Wagner about
  harmonic crystals are gratefully acknowledged.}\\[0.1cm]

\end{document}